\def\bq{\begin{equation}}
\def\eq{\end{equation}}
\def\bqa{\begin{eqnarray}}
\def\eqa{\end{eqnarray}}
\def\bqb{\begin{eqnarray*}}
\def\eqb{\end{eqnarray*}}
\documentstyle[12pt]{article}\textwidth 16cm
\def\pr#1#2#3{ Phys. Rev. ${\bf{#1}}$ (#2) #3 }
\def\prl#1#2#3{ Phys. Rev. Lett. ${\bf{#1}}$ (#2) #3 }
\def\pl#1#2#3{ Phys. Lett. ${\bf{#1}}$ (#2) #3 }
\def\prep#1#2#3{ Phys. Reports ${\bf{#1}}$ (#2) #3 }
\def\np#1#2#3{ Nucl. Phys. ${\bf{#1}}$ (#2) #3 }
\def\zp#1#2#3{ Z. Phys. ${\bf{#1}}$ (#2) #3 }

\def\ie{{\it i.e.\/}}
\def\eg{{\it e.g.\/}}

\def\etal{{\it et.al.\/}}

\def\A{ {\cal A }}
\def\B{ {\cal B }}
\def\C{ {\cal C }}
\def\R{ {\cal R}}

\global\nulldelimiterspace = 0pt

\def\hf{{1\over 2}}

\def\dd#1#2{{{d #1} \over {d #2}}}  

\def\roughly#1{\mathrel{\raise.3ex
    \hbox{$#1$\kern-.75em\lower1ex\hbox{$\sim$}}}}

\def\lam2p{\lambda^{\prime}_2}
\def\mld{m_l^2}
\def\kd{k^2}

\begin{document}
\pagenumbering{arabic}
\thispagestyle{empty}
\def\thefootnote{\fnsymbol{footnote}}
\setcounter{footnote}{1}

\begin{flushright} RAL-94-107 \\ THES-TP 94/12 \\
September  1994 \end{flushright}
\vspace{2cm}
\begin{center}
{\bf\large Searching for Exotic Mesons in $e^+e^-$
annihilation at $DA{\Phi}NE$} \footnote{ Work supported by the
European Community Human Mobility program ``Eurodafne",
Contract CHRX-CT92-0026}
\vspace{1cm} \\
{\bf\large  F.E. Close}\\
Rutherford Appleton Laboratory,\\
Chilton, Didcot, Oxon,
OX11 OQX, England,
\vspace{0.5cm}\\
and
\vspace{0.5cm}  \\
 {\bf\large  G.J. Gounaris } \\
Department of Theoretical Physics, University of Thessaloniki,\\
Gr-54006, Thessaloniki, Greece.
\end{center}
\vspace {1cm}
\begin{abstract}
\noindent
$DA{\Phi}NE$ should be useful for testing the nature of the
vector mesons directly produced in $e^-e^+$ annihilation,
as well as the nature of the $C=+$ states produced in the
radiative decays of these vector mesons. In the $DA{\Phi}NE$
energy range, these latter states may have spin up to two and any parity.
Provided the background can be handled and the spin of the
produced state is known, the study of the angular distributions in these
decays can measure the parity of the produced meson as well as
the ratios of the independent production helicity amplitudes. These ratios
provide sensitive tests for quark model classification of the states,
in particular the enigmatic $f_1(1420)$. Thus, they can be used to
indicate whether the mesons involved
are consistent with the usual
$q \bar q$  interpretation,  or whether other interpretations
like \eg\,hybrid, glueball or multiquark, are more
favorable for some of them.
\end{abstract}
\vspace{3cm}

\def\thefootnote{\arabic{footnote}}
\setcounter{footnote}{0}
\clearpage

The $DA{\Phi}NE$ \cite{dafne} $\phi$ factory will eventually access
electron positron annihilation
 at energies  above the $\phi(1020)$ mass, reaching to 1.5 or even
possibly 2 GeV in the c.m.  This will enable the
study of the dynamical
nature of mesons lying
in a mass region that is
central to the quest for gluonic
hadrons \cite{PDG}, as well as enabling the test of radical
ideas on the nature
of confinement \cite{VG,CDGKR,WI,CIK}.
Thus, we may be able to identify the conventional $q\bar{q}$
mesons in the 1-2 Gev region, and to investigate the nature of certain
controversial or unconventional states that do not appear to fit
naturally in
the quark model description of hadrons \cite{PDG,Montanet}.\par

The most direct application of $DA{\Phi}NE$ is in the production of
vector mesons in
$e^-e^+$ annihilation. Its primary role is, of course, to
concentrate on the
$\phi$(1020), but $DA{\Phi}NE$  may also help to clarify
the existence and couplings of peculiar states like
$\omega_x(1100, \ \Gamma=30 \pm 24)$ \cite{Bar},
$\rho_x(1266, \ \Gamma=110 \pm 35)$ \cite{Aston},
$\omega(1440)$ and $\rho(1460)$
which have been claimed before\footnote{ For the masses
of the latter two states we follow the findings in
\cite{CD2}.}\cite{Donn,PDG,CD2}. \par

The I=0,1 vector mesons  $\omega(1440)$,
$\omega(1600)$,
$\rho(1460)$ and $\rho(1700)$,  have found their way to the PDG list
\cite{PDG} and should be considered to some extent as
(containing at least) true resonances. The latest  analysis
of their properties is given in \cite{CD2}. In the past,
they have been most simply interpreted
as  $^3S_1$ and $^3D_1$ quarkonia states \cite{CD1},
however, more recent analyses by Clegg Donnachie and Kalashnikova
\cite{CD2,AA} conclude that this interpretation is probably
inadequate. One possibility for
a consistent description of the vector  mesons in the 1.2 to 1.7 GeV
range arises if $q\bar q$-glue hybrids are being manifested,
of which $\omega(1440)$ and $\rho(1460)$ are specific examples \cite{AA}.
An alternative picture is that one does without hybrids, but
allows for mixing with $qq\bar q \bar q$ states. In this latter
case one requires
further $qq\bar q \bar q$ states, both $\rho$ and $\omega$-like,
which are relatively low lying \cite{AA}.
Donnachie, Clegg and Kalashnikova \cite{DK}
argue that the controversial $\rho_{x}(1266)$
and $\omega_{x}(1100)$  mentioned above, may be such states. Thus,
their existence is pivotal in
discriminating between the hybrid and the $qq\bar q\bar q$ picture.
Specifically, in the hybrid picture theses states have no place
whereas  they may be accommodated in the $qq\bar q \bar q $ interpretation.
Important experiments attempting at clarifying their existence
could therefore be possible  at $DA{\Phi}NE$.\par

The state
$\omega_x(1100)$ was once claimed to
have been seen in Bethe-Heitler interference \cite{Bar}.
The mass of this state was found to be $1097 \pm 19 ~MeV$,
while its total and electronic widths are given as
$\Gamma = 30 \pm 24 ~MeV $
and $\Gamma(e^-e^+)= 50-100 eV$. From this we would
conclude that in the absence of any other contributions at
the peak of the resonance
\bq
\delta R=\sigma(e^-e^+ \rightarrow \, hadrons)/
\sigma(e^-e^+ \rightarrow \mu^+\mu^-) =
0.6 \pm 0.4 \ \ \ \ . \ \
\eq
This is 1\% of the $\phi$ signal and may be separable
from the $\phi$
with good statistics and resolution.
 Hence a careful study on the high mass side of
the $\phi(1020)$
can decide whether $\omega_x(1100)$ and/or $\rho_x(1266)$
really exist. \par

$DA{\Phi}NE$ may also be used to study the nature of $C=+$
states. The most fruitful way to do so is by measuring at the
peak of $\rho(1460)$, whose mass is most recently determined to
be $1.463 \pm 0.025~GeV$, while the  electronic
and hadronic widths are $\Gamma(e^-e^+) = 1.6~-~ 3.4~keV$
and $\Gamma_{tot} = 311 \pm 62 MeV$ \cite{CD2}. These results imply
 an increase of the $e^-e^+$ annihilation cross
section at its peak corresponding to
\bq
\delta R=\sigma(e^-e^+ \rightarrow \rho(1460) \rightarrow
\, hadrons)/\sigma(e^-e^+ \rightarrow \mu^+\mu^-) \sim 1.3
\ \ \ \ \ \ .\ \ \ \
\eq
With a luminosity of $10^{32}cm^{-2}s^{-1}$, some months of
running at the higher energies
at $DA{\Phi}NE$ will enable the study of radiative
decays of vector mesons like \eg\@ $\rho(1460)$
into $C=+$ states. Several of these states, having spin up to
two and either parity, are particularly interesting.To motivate
our more detailed discussion we present first some illustrative examples.
\par

The easiest interpretation of these
measurements arises if
$\rho(1460)$ is a $^3S_1$ quarkonium state, (as expected in the Godfrey
and Isgur picture of the quark model spectrum \cite{gi}), which could
be tested by looking at its radiative decay to the well
established $^3P_{2,1}$ states $f_2(1270)$, $a_2(1320)$,
$f_1(1285)$ and $a_1(1260)$. In such a case the independent helicity
amplitudes satisfy certain constraints (to be discussed later, see
eqs (16) and (23)): if any of these relations is violated
then we would conclude that $\rho(1460)$
is not simply an ordinary $^3S_1$ quarkonium. Further useful
knowledge may be subsequently acquired by searching for the $\rho(1460)$
radiative decay to the controversial
$f_2(1430)$, which has been claimed in $J/\Psi$ radiative decays,
(see p.1486 in ref.\cite{PDG})
and compare the ratios of its
various helicity production amplitudes to those for the well
established $^3P_2$ quarkonium state $f_2(1270)$ \cite{PDG}.
The ~interpretation of this latter measurement would of course
depend on whether the previously mentioned measurement supports
the quarkonium assignment of  $\rho(1460)$. \par

As another notable example we could measure the single amplitude
determining the production of the $0^{++}$ states $f_0(975)$
or  $a_0(980)$. This way we can test whether these mesons are
the $^3P_0$ analogs of the well established $^3P_2$ quarkonium
state $f_2(1270)$, or whether one of them at least is exotic
\cite{CDGKR,WI,CIK}. The peculiarities of  the $1^{++}$
sector can also be studied
at DA$\Phi$NE. The interest here stems from the fact that the PDG
list already includes ten axial mesons in this region, not all of
which can be quarkonia. The odd one
out appears to be the $f_1(1420)$. This state has also been seen in
$\gamma-\gamma^*$   and may be a $KK^*$ molecule or even a
$1^{++}$ hybrid state \cite{f1420,Barnes}. There exist also some
questions concerning its parity so that the quark model exotic
$J^{PC} = 1^{-+}$ is not fully excluded \cite{PDG}. \par

In the present paper, we consider
ways of elucidating this possibility by looking at the
exclusive production of such states in the radiative decays of
$\rho$(1460) or in the $e^-e^+$ continuum. We always assume that
we know the spin $J$ of the produced state and that we somehow
understand the background. These processes
enable us to measure the ratios of the helicity amplitudes
determining the production of these states, and thereby test
their parity. Moreover, since the quark
model makes unambiguous predictions for these ratios, very strong
constraints on possible deviations
from the quarkonium-like structure could be imposed. \par

The radiative decay for
\bq
e^-e^+ \to 1^{--} \to J^{P+} ~~ \gamma \ \ \ \ \ ,\ \ \ \ \
\eq
(where $J^{P+}$ denotes any $C=+$ state with $P = \pm$),  is described by
\bq
\dd{\sigma (e^-e^+ \to 1^{--} \to
J^{P+} \gamma )}{\cos \theta} =
{\sum_{m_1=\pm 1}}\sum_{\lambda_1
\lambda_2}\left [ d^1_{m1,\lambda_1-\lambda_2}(\theta)
 ~ |\langle \lambda_1 \lambda_2 | T^1(0)|m_1 \rangle |
  \right ]^2 \ \ . \
\eq
Here $m_1$ measures the projection of the spin of the initial
$1^{--}$ state along the
$e^-$ beam taken as the $z$ axis in the $e^-e^+$ c.m.\@
frame. Its only allowed values
are $m_1=\pm 1$.  The production angle of the
$J^{P+}$ state with respect the same $e^-$ axis is denoted by $\theta$,
while $\lambda_1$ and $\lambda_2$ describe the helicities of the
$J^{P+}$ state and the photon respectively.
The various partial wave helicity
amplitudes for the radiative decay, (to lowest order in $\alpha$),
of the initially formed $1^{--}$ state, are  given by
$\langle \lambda_1 \lambda_2 | T^1(0)|m_1 \rangle$,
where the argument of $T^1$ emphasizes that the final photon in
process (3) is on its mass shell; \ie\@ the photon momentum
$k_\mu$ satisfies $k^2=0$.
These amplitudes depend on the dynamics
responsible for the existence of the
initial $1^{--}$ and the final $J^{P+}$ states. For $J\geq 1$,
where more than one amplitude contributes, the $\theta $
distribution is determined  by
the ratios of these amplitudes and provides therefore
a strong test of the quark model. \par

More information on these ratios  may be obtained
by looking also at the emission of a virtual photon decaying
to a $l^-l^+$ pair in a process such as
\bq
e^-e^+~\to 1^{--} ~\to ~J^{P} ~~ \gamma^* ~
\to ~J^{P+} ~~ (l^-l^+) \ \ \ \
\ \ \ . \ \ \ \ \ \ \
\eq
Now the final state distribution depends also
on the angle $\varphi$, defined as the angle between the
$J^{P+}$ production plane and
the decay plane of $\gamma ^* \to l^-l^+$. By definition
$0\leq \varphi \leq \pi/2$. The final state distribution depends also on
 the
angle $\theta$ and on the squared momentum
$k^2$ of $\gamma^*$ (where $\kd > 4 \mld $) such that
\bqa
 \frac{d \sigma (e^-e^+ \to 1^{--} \to
J^{P+} \gamma^* \to J^{P+} (l^-l^+))}{d k^2 d \cos \theta d\varphi }
 =\frac{\alpha}{16\pi^2}   \frac{1}{k^4}\left (1-\frac{4m_l^2}{k^2}
\right)^{1/2} \ \ \ \ \ \ \ \ \ \ \ \ \ \ \ \ \ \nonumber \\[0.5cm]
\cdot  { \sum_{m_1=\pm 1}}\sum_{\lambda_1
\lambda_2 \lam2p } d^1_{m1, \lambda_1- \lambda_2 }(\theta)
d^1_{m1,\lambda_1 -\lam2p }(\theta) L_{\lam2p \lambda_2}
\langle \lambda_1 \lambda_2 | T^1(k^2)|m_1 \rangle
\langle \lambda_1 \lam2p | T^1(k^2)|m_1 \rangle ^*
  \ \ \ \ \ , \ \ \ \
\eqa\\
where again $\lambda_1$ is the $J^{P+}$ helicity, $\lambda_2$
or $\lam2p$ denote the helicity of $\gamma^*$, and
$L_{\lam2p \lambda_2}$ is the $\gamma ^*$ density matrix.
Denoting by $m_l$ the mass of the final leptons and
integrating out the $l^-$ polar angle in the  $\gamma ^*$
decay plane, we find
\bq
L_{\lam2p \lambda_2}~=~\frac{8}{3} \left (
\begin{array}{ccc}
k^2+2m_l^2 &    0   & \hf (k^2-4m_l^2)\exp{i2\varphi} \\[.3cm]
 0           &   k^2+2m_l^2  &  0 \\[.3cm]
 \hf (k^2-4m_l^2)\exp{-i2\varphi}          & 0  &  k^2+2m_l^2 \\[.3cm]
\end{array}
\right )
\ \  , \
\eq
where the rows and columns correspond to $\gamma ^*$
helicities $+1,0,-1$.
Of course, process (5) is suppressed with respect to (3) by an
extra power of $\alpha$ and it is useful only so far as the
$\varphi$ distribution is needed.\par

Since in $DA{\Phi}NE$
 $\kd$ cannot be very large,  $\gamma^*$ is very close to its
mass shell and behaves almost like a real photon. Therefore
we can neglect all amplitudes involving
a longitudinal $\gamma^*$, and conclude therefore that the same
kind of helicity amplitudes,
involving only transverse photons, contribute in both (3,4)
and (5,6). After integrating over $\varphi$ in (6,7),
$L_{\lam2p \lambda_2}$ becomes essentially the
unit matrix, which means that the
$\theta$ distribution for on and off-shell photons is given by
the same expression in terms of the helicity amplitudes; compare
(4, 6). Parity and time inversion invariance for such
amplitudes imply respectively
\bqa
\langle \lambda_1 \lambda_2 | T^1(k^2)|m_1 \rangle~ =~
P\, (-1)^J \langle -\lambda_1 ~-\lambda_2 | T^1(k^2)|m_1 \rangle
\ \ \ \ , \ \ \
\\[0.5cm]
\langle \lambda_1 \lambda_2 | T^1(k^2)|m_1 \rangle ~=~-
{}~\langle \lambda_1 \lambda_2 | T^1(k^2)|-m_1 \rangle ^*
\ \ \ \ \ \ \ \ \ \ \ \ \ \ \ \ . \ \
\eqa\par

In the following we apply (4, 6, 8, 9) to the production of
resonances with specific spin $J$ and parity $P=\pm 1$.\par

\underline{J=0 States}.
 We start from the radiative production of a $J=0$
state with parity $P$, for which there is only one independent
helicity amplitude taken to be
\bq
\A(\kd)~=~\langle  \lambda_2=+ \, | T^1(\kd)|m_1=+ \, \rangle
\ \ \ \ \ . \ \ \ \ \
\eq
This leads to
\bq
\frac{d\sigma}{d\cos \theta} ~\sim ~ (1+ \cos^2 \theta)|\A(0)|^2 \
\ \ \ \ \ \ \ ,\ \ \ \ \
\eq
for process (3) and
\bq
\frac{d \sigma}{d k^2 d \varphi}~\sim ~
\frac{\alpha}{k^2}\left (1-\frac{4m_l^2}{k^2}\right)^{1/2}
|\A(\kd)|^2 \Bigg \{\left(1+\frac{2\mld}{\kd}\right)+
\frac{P}{4}\, \left(1-\frac{4\mld}{\kd}
\right ) \cos (2\varphi) \Bigg\}
\ \ \ \ \ \ \  , \ \ \ \ \ \eq
for (5). Eq (12) indicates that the sign of the $\cos (2\varphi)$
coefficient discriminates between the two different parities
($P = \pm$) of
the produced $J=0$ state; thus for example the $\cos (2\varphi)$ coefficient
for $f_0(975)$ and $a_0(980)$ should be opposite to
those of $\eta$ and $\eta'(960)$ \cite{GN}; as we will see below
it is also true for higher
spin resonances. The overall production rate of $0^{++}$ mesons
is described by an E1 transition and therefore is described by one
amplitude only. Thus in the quark model the
production of the $^3P_0$ quarkonium state is fully determined by
the radiative production of the well established $^3P_2$
quarkonia $f_2(1270)$ and $a_2(1320)$; see below\cite{FECbook,Rosner}.
 Therefore, means
of testing whether $f_0(975)$ or  $a_0(980)$ are the $^3P_0$ analogs
of the well established $^3P_2~ f_2(1270)$, $a_2(1320)$
are supplied. The same process may also be used to study the
quark content of the candidates for radially excited $0^{-+}$ states
namely $\pi(1300)$, $\eta(1295)$ \cite{PDG} and $\eta(1400)$ claimed by
MARK III in the $K\bar K \pi$ and $\eta \pi \pi$ modes
\cite{Bai}.\par

\underline{J=1 States}.
We next consider the $J=1$ case. This has the extra
interest that a quarkonium structure for the final
state meson is only possible in the $1^{++}$ configuration, and not the
$1^{-+}$ one. Parity and time inversion invariance imply that
there exist only two independent amplitudes which we take to be
\bq
 \B_1(\kd)~=~ \langle ++ | T^1(\kd)| + \rangle \ \ \ \ \ \ , \ \ \ \ \
\B_0(\kd)~=~ \langle 0 + | T^1(\kd)|+ \rangle \ \
 \ \ . \ \ \ \
\eq
Substituting in (4,6), using (8,9), we get
\bq
\frac{d\sigma}{d\cos \theta} ~\sim ~ (2|\B_1(0)|^2 + |\B_0(0)|^2 )
\Bigg [1 ~+~ \frac{(|\B_0|^2-2|\B_1|^2)}{(2|\B_1(0)|^2 + |\B_0(0)|^2)}
{}~\cos^2(\theta) \Bigg ]
\ \ \ \ \ \ \ ,\ \ \ \ \
\eq
for process (3) and
\bqa
\frac{d \sigma}{d \kd d \varphi} \sim
\frac{\alpha}{k^2}\left (1-\frac{4m_l^2}{k^2}\right)^{1/2} \Bigg
\{ \left(1+\frac{2\mld}{\kd}\right)(|\B_1(\kd)|^2 +|\B_0(\kd)|^2 )
-  \nonumber \\[0.4cm]
- \frac{P}{4} |\B_0(\kd)|^2 \left(1-\frac{4\mld}{\kd}
\right ) \cos (2\varphi) \Bigg\}
 \  \ \ \ \  , \ \ \ \
\eqa
for (5). We note in (15) that the sign of the $\cos (2\varphi)$
coefficient determines again the parity of the produced state.
{\bf Thus, provided $J=1$ for the produced state is established,
a positive sign for the $cos(2\phi)$ term will be
an unambiguous signature that an
exotic $1^{-+}$ state is formed}. Taking into account
the fact that the $^3P_1~~1^{++}$ nonet is
already complete and that $f_1(1420)$ appears as a tenth state,
it will be very interesting to use (15) in order to measure
its parity and discriminate
between a vector or axial vector assignment
\cite{PDG,f1420}.  The fact that $f_1(1420)$ has already been
seen in $\gamma \gamma^* $ fusion argues in favor of the
feasibility of such a search \cite{f1420}.  \par

Leaving this aside, we now consider the production of an $1^{++}$ state
through process (3). According to the quark model, if this state
is a $^3P_1$ quarkonium and the initial decaying one is
an ordinary $^3S_1 ~~ q
\bar q$ vector meson,  then the amplitudes are determined mainly
by  E1 and M2 transitions
which, in an obvious notation, imply \cite{FECbook,CL}
\bq
\B_1(0)~=~\sqrt{3}\Big (E+\frac{M}{2} \Big )\ \ \ \ \ \ \
, \ \ \ \ \ \ \ \ \B_0(0)~=~\sqrt{3}\Big (E-\frac{M}{2} \Big )
\ \ \ \ \ \ , \ \ \ \
\eq
for real photons in the final state; (compare (13)).
Let us now specialize to the case that the initial
$e^-e^+$ energy is around the $\rho(1460)$ mass, and that we
are interested in studying the structure of a state in the
$f_1(1420)$ region.
In such a case the momentum of the outgoing photon is about 40
MeV, which, according to the quark model implies strong
E1 dominance and $\B_1/\B_0\sim 1$.
For such a case (14) gives $d\sigma/\cos \theta \sim
[1 - \cos^2(\theta)/3 ]$ \cite{FECbook,Rosner}.
For process (5) on the other hand, where $\kd$ is non-vanishing,
the ratio $\B_1(\kd)/\B_0(\kd)$ will grow dramatically with $\kd$
 \cite{CG}. Such a rise
has already been seen in baryons and has important implications for
the spin dependent electroproduction \cite{FEC1974}.
In the DA$\Phi$NE kinematical region however, the final
$\gamma^*$ is only slightly off shell, implying again
$\B_1(\kd)/\B_0(\kd)  \simeq 1$ and a small $M/E$ value.\par

Thus, the basic quark model prediction which can be tested through
(14) and (15), is that the
production of a true $^3P_1~~1^{++}$ quarkonium state through radiative
decay of a $^3S_1~~q\bar q$ vector meson, is characterized by
$\B_1/\B_0 \simeq 1$ in the DA$\Phi$NE kinematical region.
This prediction is quite strong and safe. The main problem is
background. If this is overcome and the $1^{++}$ quantum numbers
of the produced state are
established, we can conclude that any indication from (14-16)
implying that $|M/E|$ is large and thus $\B_1/\B_0$ very different
from $+1$,
would either mean that the initial state is
$^3D_1$ (which is hard to imagine in this mass region, see ref
\cite{gi}),  or that
at least one of the participating mesons is exotic \cite{FECbook,Rosner,CG}.
 \par

\underline{J=2 States}.
As a final example, we consider the production of a $2^{P+}$ state
through process (3) and (5). The number of independent
amplitudes after parity and time inversion invariance is taken
into account, become now three. For these amplitudes we take
\bq
\C_2(\kd) = \langle 2~1 |T^1(\kd) |+\rangle \  \ ,  \ \
\C_1(\kd) = \langle 1~1 |T^1(\kd) |+\rangle \ \  ,  \ \
\C_0(\kd) = \langle 0~1 |T^1(\kd) |+\rangle \ \  .   \
\eq
Substituting in (4,6), using (8,9) we get in this case
\bq
\frac{d\sigma}{d\cos \theta} ~\sim ~
(|\C_0(0)|^2 + |\C_2(0)|^2 + 2|\C_1(0)|^2 )
[1 + \R_\theta \, \cos^2(\theta)  ]
\ \ \ \ \ \ \ ,\ \ \ \ \
\eq
with
\bq
\R_\theta =  \frac{(|\C_0(0)|^2 + |\C_2(0)|^2 - 2|\C_1(0)|^2 )}
{(|\C_0(0)|^2 + |\C_2(0)|^2 + 2|\C_1(0)|^2 )} \ \ \ \ \  \ \ \
\ \ \
\eq
for process (3), and
\bqa
\frac{d \sigma}{d \kd d \varphi} \sim
\frac{\alpha}{k^2}\left (1-\frac{4m_l^2}{k^2}\right)^{1/2}
(|\C_0(\kd)|^2 +|\C_1(\kd)|^2 +|\C_2(\kd)|^2 )
\Bigg \{ \left(1+\frac{2\mld}{\kd}\right)
+  \nonumber \\[0.4cm]
P \, \R_\varphi \left(1-\frac{4\mld}{\kd}
\right ) \cos (2\varphi) \Bigg\}
 \  \ \ \ \  , \ \ \ \
\eqa
with
\bq
\R_\varphi =\frac{|\C_0|^2}{4 (|\C_0(\kd)|^2 +|\C_1(\kd)|^2
+|\C_2(\kd)|^2 )}\ \ \ \ \ \ \ \
\eq
for (5). Again the sign of the $\cos (2\varphi)$ distribution
determines the parity of the produced $2^{P+}$ state. The
results (19, 21) imply also
\bq
-1 \leq \R_\theta \leq 1 ~~~~~~ , ~~~~~~
0 \leq \R_\varphi \leq \, \frac{1}{4} ~~~~~~~~~~~ ,~~~
\eq
as a consequence of parity and time inversion invariance and the
approximation to neglect amplitudes involving longitudinal $\gamma ^*$ in
the DA$\Phi$NE kinematical region for process (5).\par

In the remaining we restrict to the production of a $2^{++}$
which may be more interesting for DA$\Phi$NE. If this state
happens to be a quarkonium $^3P_2$, then standard quark
model considerations imply \cite{FECbook,Rosner}
\bq
\C_0=E -\frac{3M}{2} ~~~ , ~~~
\C_1= \sqrt{3}\left (E-\frac{M}{2} \right )~~~ , ~~~
\C_2 =\sqrt{6} \left (E -\frac{M}{2} \right ) ~~~ ,~~
\eq
where the $\kd$ dependence is suppressed. Combining (23) with
(19) and (21) we get
\bq
-\frac{3}{11}\, \leq \R_\theta \leq 1 ~~~~ , ~~~~
0 \leq \R_\varphi \leq \, \frac{3}{20} ~~~~ , ~~~
\eq
which should be satisfied if the produced state is a $^3P_2$
quarkonium. This quarkonium prediction can be made even more
restrictive if E1 dominance for the amplitudes in (23) is taken
into account, which gives
\bq
\R_\theta =\, \frac{1}{13} ~~~~~, ~~~~~~ \R_\varphi =\,
\frac{1}{40} ~~~~~ .~~~
\eq
 Provided that the production of
a $2^{++}$ state is established, any violation of (24)
will be a proof that an exotic $2^{++}$ state has been
identified. The same conclusion could also be drawn if a
sufficiently strong violation of (25) is observed , implying an
unacceptably large value for $M/E$.\par

Apart from a comment in the discussion after (12), up to now
we have not used the less general quark model
predictions relating
the amplitudes for producing $P=C=+$ mesons with different $J$.
If such relations are used, then connections between the $\A$,
$\B_i$ and $\C_j$ amplitudes are found. These connections stem
from the E1 dominance found in the quark model for the $\B_i$ and
$\C_j$ amplitudes, which combined with the fact that the $\A$
amplitude for the $0^{++}$ production is purely E1 induced,
leads to (modulo phase space effects)
\bq
\sigma (e^-e^+ \to ~  V \to ~ \gamma(E1) ~+~ ^3P_{0,1,2}) = 1:3:5
\ \ \ \ \ \ \  . \ \ \ \ \
\eq
These relations may be used to verify the tacit assumption that the
axial meson $f_1(1285)$  is the
$^3P_1$ quarkonium partner of the established $^3P_2
{}~~f_2(1270)$, and to explore the nature of  $f_1(1420)$.
Furthermore, we can also study whether either of $f_0(975)$ and $a_0(980)$
is a  conventional $^3P_0$ quarkonium partner of $f_2(1270)$,
or whether they have some other dynamical structure.
The $f_0(975)$ is particularly interesting in this regard.
According to (26), if it is the $^3P_0$ partner of
$f_2(1270)$, its production
strength is fixed. Conversely, if different production rates
are seen, it may be
possible to deduce whether $f_0(975)$ is a ``small" (e.g.\@ \cite{CDGKR}) or
``large" (e.g.\@ \cite{WI}) system; since the
E1 transition amplitude is in proportion to
($\Sigma_i e_i r_i$), where $e_i$ and $r_i$ are the
charge and vector displacement of the charged constituents
(see  \cite{CIK}).\par

We have shown in the present paper that provided the background
can be handled, the radiative production of $C=+$ states in
$e^-e^+$ annihilation at DA$\Phi$NE can be very helpful in
indicating whether the mesons involved
are consistent with the usual
$q \bar q$  interpretation,  or whether other interpretations
such as \eg\, hybrid, glueball or multiquark, are more
favorable for some of them.

\underline{Acknowledgements}:
GJG would like to thank the RAL Theory Group for the hospitality
offered to him during his visit Rutherford Appleton Laboratory
where most of this work
was done.

\newpage

\end{document}